\documentclass[manuscript]{aastex_mod}
\usepackage{amsmath}
\usepackage{amssymb}
\usepackage{graphicx}
\usepackage{amssymb}
\usepackage{epsfig}
\usepackage[sort&compress]{natbib}

\shorttitle{Co-colonization with different MRSA strains }
\shortauthors{D'Agata et al.}

\begin{document}

\begin{center}
\textbf{Rapid emergence of co-colonization with community-acquired and\\
hospital-acquired methicillin-resistant \textit{Staphylococcus aureus} strains in\\
 the hospital setting }

\

Erika M. C. D'Agata$^1$, Glenn F. Webb$^2$, and Joanna Pressley$^2$\\

\

$^1$Division of Infectious Diseases, Beth Israel Deaconess Medical Center, Harvard Medical School, Boston, MA 02215, $^2$Department of Mathematics, Vanderbilt University, Nashville, TN 37240\\
\end{center}

\noindent Keywords: methicillin resistance, \textit{Staphylococcus aureus}, community, hospital, co-colonization\\

\noindent Running title: co-colonization with different MRSA strains \\

\noindent Correspondence to:\\
	Erika D'Agata MD MPH\\
    Beth Israel Deaconess Medical Center, Division Infectious Diseases\\
	330 Brookline Ave, East Campus Mailstop SL-435G\\
	Boston, MA 02215\\
	e-mail address: edagata@bidmc.harvard.edu\\
	phone (617) 667-8127; fax (617) 667-7251


\begin{abstract}
\noindent\textbf{Background:} Community-acquired methicillin-resistant \textit{Staphylococcus aureus} (CA-MRSA), a novel strain of MRSA, has recently emerged and rapidly spread in the community.  Invasion into the hospital setting with replacement of the hospital-acquired MRSA (HA-MRSA) has also been documented.  Co-colonization with both CA-MRSA and HA-MRSA would have important clinical implications given differences in antimicrobial susceptibility profiles and the potential for exchange of genetic information.\\
\textbf{Methods:} A deterministic mathematical model was developed to characterize the transmission dynamics of HA-MRSA and CA-MRSA in the hospital setting and to quantify the emergence of co-colonization with both strains.\\
\textbf{Results:} The model analysis shows that the state of co-colonization becomes endemic over time and that there is no competitive exclusion of either strain.  Increasing the length of stay or rate of hospital entry among patients colonized with CA-MRSA leads to a rapid increase in the co-colonized state.  Compared to MRSA decolonization strategy, improving hand hygiene compliance has the greatest impact on decreasing the prevalence of HA-MRSA, CA-MRSA and the co-colonized state.\\
\textbf{Conclusions:}  The model predicts that with the expanding community reservoir of CA-MRSA, the majority of hospitalized patients will become colonized with both CA-MRSA and HA-MRSA.
\end{abstract}

\section*{Introduction}
Methicillin-resistant \textit{Staphylococcus aureus} (MRSA) has been traditionally considered a hospital-acquired bacteria and is implicated in the great majority of infections acquired in the hospital \cite{1}.  The documentation of a novel MRSA strain, which has emerged in the community and has subsequently spread into the hospital, has led to a re-evaluation of the transmission dynamics of MRSA in the healthcare setting \cite{2,3}.  Several population-based surveillance studies have documented that the community-acquired MRSA (CA-MRSA) may be replacing the hospital-acquired MRSA (HA-MRSA) \cite{4,5,6}.  Mathematical models corroborate these findings and predict that there will be competitive exclusion of HA-MRSA strains by CA-MRSA over time \cite{7}.
	Previous studies have assumed that individuals can only be colonized or infected with either HA-MRSA or CA-MRSA.  Data suggest however, that individual colonization with multiple \textit{Staphylococcus aureus} strains occurs \cite{8}.  Co-colonization with multiple strains of other bacterial species has also been documented \cite{9}.  Understanding the emergence and spread of co-colonization with both CA-MRSA and HA-MRSA would have important clinical implications given differences in antimicrobial susceptibility profiles and virulence properties between these two strains \cite{10}.  Co-colonization can also result in the horizontal transfer of mobile genetic elements between strains, such as antimicrobial-resistance or virulence determinants.  These events may lead to the emergence of MRSA strains with novel biological properties.
	A mathematical model was developed to understand the emergence and spread of co-colonization with both CA-MRSA and HA-MRSA among hospitalized individuals.  This model extends a previous model characterizing the transmission dynamics of CA-MRSA into the hospital setting, which assumed that patients could only be colonized or infected with either CA- or HA-MRSA \cite{7}.  Key factors, which contribute to the spread of antibiotic-resistant bacteria and their impact on the emergence of a co-colonized state in the hospital setting, were analyzed through model simulations.  The impact of an increased influx of patients harboring CA-MRSA into the hospital was quantified using data from population-based surveillance studies, which document an expanding community reservoir of CA-MRSA.  Differences in length of stay (LOS) among patients harboring CA-MRSA were also analyzed since patients infected with CA-MRSA can present, with severe infections leading to longer LOS.  Lastly, infection control strategies aimed at limiting the spread of MRSA and their effect on the emergence of the co-colonized state were evaluated.

\section*{Methods}
\noindent\textbf{Mathematical Model}


A deterministic model was developed to characterize the transmission dynamics of HA-MRSA and CA-MRSA in the hospital setting and to quantify the emergence of co-colonization with both CA-MRSA and HA-MRSA among hospitalized patients.

\begin{figure}
\includegraphics[width=.65\textwidth]{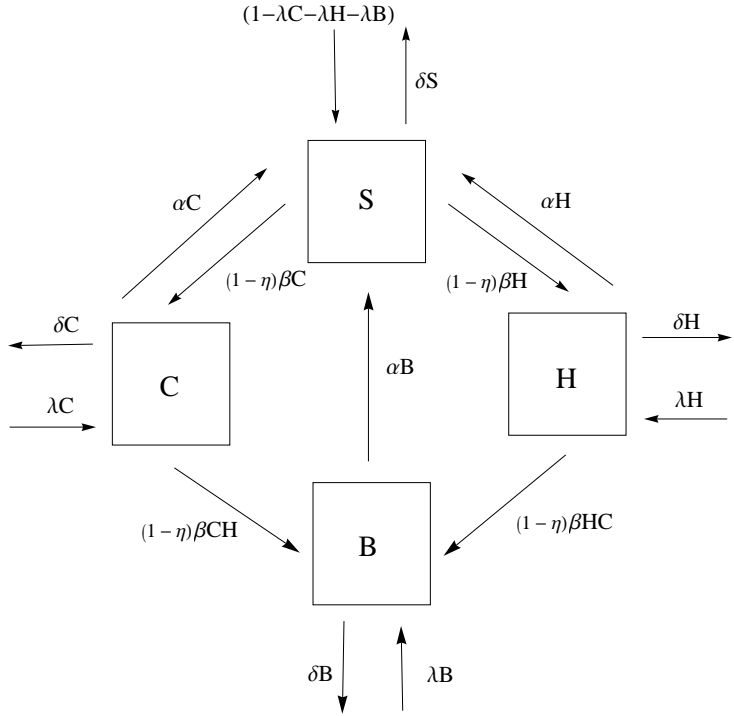}
\caption{A compartment model describing the transmission dynamics of CA-MRSA and HA-MRSA in a 400-bed hospital. The arrows and parameter values correspond to entry and exit from the 4 compartments ($S$-susceptible patients, $C$-patients colonized with CA-MRSA, $H$-patients colonized with HA-MRSA, and $B$-patients co-colonized with both strains). The percentages of patients admitted colonized with CA-MRSA, colonized with HA-MRSA, or colonized with both strains are expressed as $100\lambda_C$, $100\lambda_H$, and $100\lambda_B$, respectively. Discharge and death rates from the compartments are expressed as follows: $\delta_S$, $\delta_C$, $\delta_H$, and $\delta_B$ for susceptible patients, patients colonized with CA-MRSA, patients colonized with HA-MRSA, and patients co-colonized with both strains, respectively (with mean length of stays defined as $1/\delta_S$, $1/\delta_C$, $1/\delta_H$, and $1/\delta_B$). The colonization rates of susceptible patients to the CA-MRSA compartment  is $(1-\eta)\beta_C$ and to the HA-MRSA compartment is $(1-\eta)\beta_H$. The co-colonization rate from $C$ to the co-colonized compartment ($B$) is $(1-\eta)\beta_{CH}$ and from $H$ to $B$ is $(1-\eta)\beta_{HC}$, where 100$\eta$ signifies the percentage of hand-hygiene compliance (where $\eta=0$ corresponds to 0\% compliance and $\eta=1$ corresponds to 100\% compliance). The rates of decolonization of patients with CA-MRSA, HA-MRSA, or both strains are given by $\alpha_C$, $\alpha_H$, and $\alpha_B$, respectively.}
\label{fig1}
\end{figure}

	Individuals in the hospital are in four exclusive states: susceptible ($S$), colonized with either CA-MRSA ($C$), HA-MRSA ($H$) or both CA- and HA-MRSA ($B$). The infected state is not included.  Patients enter the hospital as $S$, $C$ or $H$.  To understand the emergence of the co-colonized state, the model assumes that there is no co-colonization at baseline and that patients do not enter the hospital in the $B$ state.  Patients leave the hospital via death or hospital discharge in all four states.  Through contact with a contaminated healthcare workers (HCW), susceptible patients becomes colonized with either CA-MRSA at a rate $(1-\eta)\beta_C$ or HA-MRSA at a rate of rate $(1-\eta)\beta_H$. Here, $\eta$ signifies compliance with hand hygiene measures with $0\leq\eta\leq1$, $\eta=0$ corresponds to no compliance and $\eta=1$ corresponds to perfect compliance. Transmission rates of CA-MRSA and HA-MRSA are given by $\beta_C$ and $\beta_H$ respectively. Once in the $C$ or $H$ state, patients can transition to the co-colonization state, $B$, through contact with a HCW, contaminated with either HA-MRSA or CA-MRSA, with rates $(1-\eta)\beta_{CH}$ or $(1-\eta)\beta_{HC}$, respectively (see figure \ref{fig1} and appendix for model details).

The LOS among CA-MRSA colonized patients was assumed to equal to the LOS of susceptible patients in the baseline model. The LOS for the co-colonized compartment starts after acquisition of the second strain and, as a simplification, is set equal to the larger of the LOS for patients colonized with CA-MRSA or the LOS for patients colonized with HA-MRSA. Since patients colonized with MRSA can develop an infection during their hospitalization, thereby substantially prolonging their LOS, simulations were performed to quantify the impact of an increasing LOS on the transmission dynamics of HA-MRSA, CA-MRSA and the emergence of the co-colonization state.  Simulations were also performed to determine the impact of an increase in the percent of patients entering the hospital already colonized with CA-MRSA.  Model simulations evaluating the impact of hand-hygiene and decolonization of MRSA colonized patients, two key infection control strategies, were also performed.  Since several different decolonization strategies are available with varying efficacy, the decolonization parameters of patients colonized with CA-MRSA ($\alpha_C$),  HA-MRSA ($\alpha_H$), or co-colonized ($\alpha_B$) ranged from 0\% efficacy (no decolonization) to 100\% efficacy.

The mathematical model assumes that the likelihood of transmission of CA-MRSA and HA-MRSA are equal ($\beta_C = \beta_H$).  In vitro data suggest that the growth rate of CA-MRSA is faster than that of HA-MRSA for certain CA-MRSA strains.  This biological difference may allow CA-MRSA to have an advantage towards colonization and subsequent transmission compared to HA-MRSA.  To understand the implications of a greater transmission potential among CA-MRSA strains, the baseline model and above simulations were re-analyzed with $\beta_C > \beta_H$ \cite{11,12}.

Parameter estimates were obtained from infection control data, microbiology data, and patient data from a 400-bed tertiary care hospital with approximately 25,000 admissions per year.  Estimates were also obtained from published studies focusing on the epidemiology of CA-MRSA or HA-MRSA (Table 1) \cite{5,11,12,13,14,15,16}.




\section*{Results}
\noindent\textbf{Transmission dynamics of HA-MRSA and CA-MRSA and co-colonization}
\noindent\textbf{\textit{Baseline model}}

\begin{figure}
\includegraphics[width=.75\textwidth]{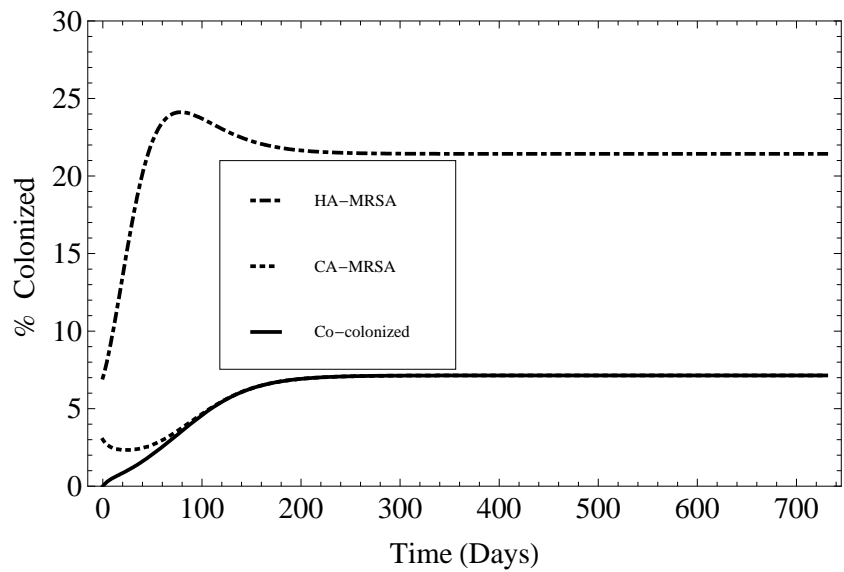}
\caption{Time evolution, over two years, of the percentage of patients colonized with HA-MRSA (dashed), CA-MRSA (dotted) or both (solid). Length of stay for HA-MRSA and both is 7 days, and 5 days for CA-MRSA. Initially the patient population consists of 90\% susceptible patients, 3\% of patients colonized with CA-MRSA, 7\% of patients colonized with HA-MRSA, and 0\% co-colonized.\label{fig2}}
\end{figure}

A baseline model was developed to quantify the prevalence of patients colonized with HA-MRSA, CA-MRSA, or both strains over time.  To understand the underlying transmission dynamics of HA- and CA-MRSA and the emergence of co-colonization with both strains, this baseline model assumes that there is no entry of patients who are already colonized with MRSA into the hospital (all entering patients are susceptible).  Analysis shows that when the basic reproduction numbers satisfy $R0^H>R0^C >1$ (see appendix), the state of co-colonization with both HA-MRSA and CA-MRSA becomes endemic over time. Assuming $\beta = \beta_C = \beta_H = \beta_{HC} = \beta_{CH}$ and $\alpha = \alpha_C= \alpha_H = \alpha_B$, the model analysis also demonstrates that there is no competitive exclusion of either strain, when both $R0^C >1$ and $R0^H > 1$.  Over time, the prevalence of patients colonized with HA-MRSA exceeds the prevalence of patients colonized with CA-MRSA since $R0^H >R0^C$.  The greater $R0^H$ value for HA-MRSA compared to the $R0^C$ value for CA-MRSA reflects the longer LOS among HA-MRSA patients, which results in greater opportunities for HA-MRSA transmission (figure \ref{fig2}).

\subsection*{Simulation 1: increased transmission of CA-MRSA and HA-MRSA }
\begin{figure}
\includegraphics[width=0.75\textwidth]{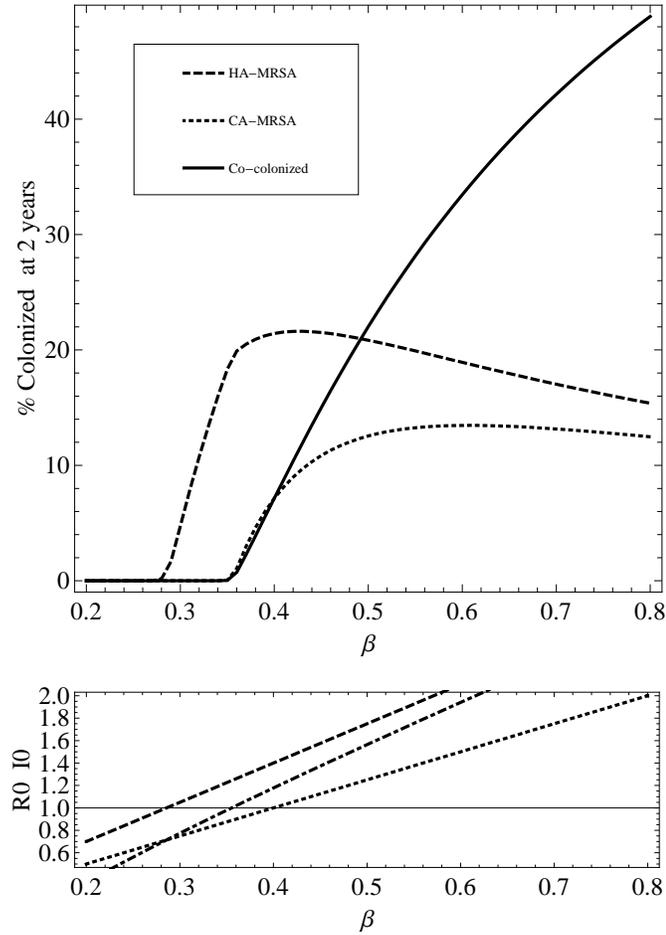}
\caption{Top - percentage of patients colonized with HA-MRSA (dashed), CA-MRSA (dotted), and both (solid) after 2 years, as transmission ($\beta=\beta_C=\beta_H=\beta_{CH}=\beta_{HC}$) is increased. Bottom - $R0^H$ (dashed) and $R0^C$ (dotted) as transmission is increased. The invasion reproduction number $I0^{Eh}$ is plotted with the dash-dotted line (see Appendix for description).}
\label{fig3}
\end{figure}

Increasing patient-to-patient transmission through patient contact with HCW contaminated with either HA-MRSA or CA-MRSA results in a substantial increase in the percent of co-colonized patients.  As transmission increases, more patients are colonized with MRSA and less are susceptible, and therefore more patients that are colonized with individual strains become co-colonized.  Above a threshold value of $\beta$, the percent of patients co-colonized with both strains exceeds those colonized with either CA-MRSA or HA-MRSA (figure \ref{fig3}).

\subsection*{Simulation 2: increasing the influx of patients colonized with CA-MRSA into the hospital and their LOS}
\begin{figure}
\includegraphics[width=1\textwidth]{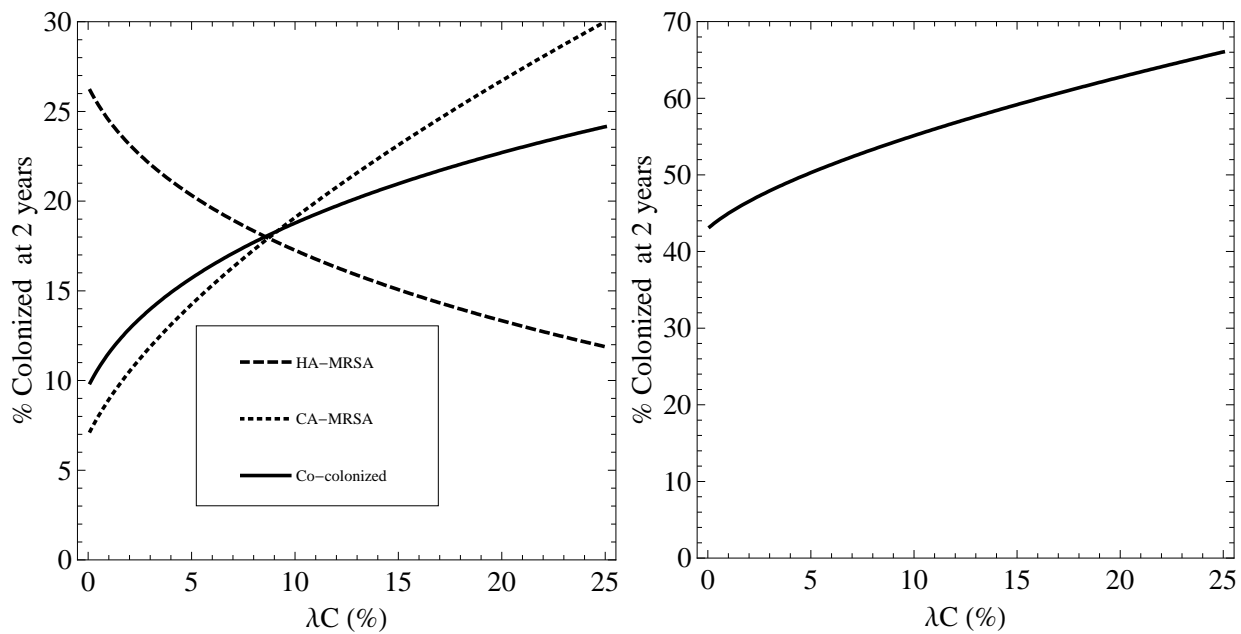}
\caption{Left - percentage of patients colonized with HA-MRSA (dashed), CA-MRSA (dotted), and both (solid) after two years, as the percentage of patients entering the hospital already colonized with CA-MRSA ($\lambda_C$) is increased. Right - the total percentage of patients colonized with MRSA after 2 years  as $\lambda_C$ is increased. In both figures, $100\lambda_H$ = 7\% and $100\lambda_B$ = 0\%.}
\label{fig4}
\end{figure}

\begin{figure}
\includegraphics[width=1\textwidth]{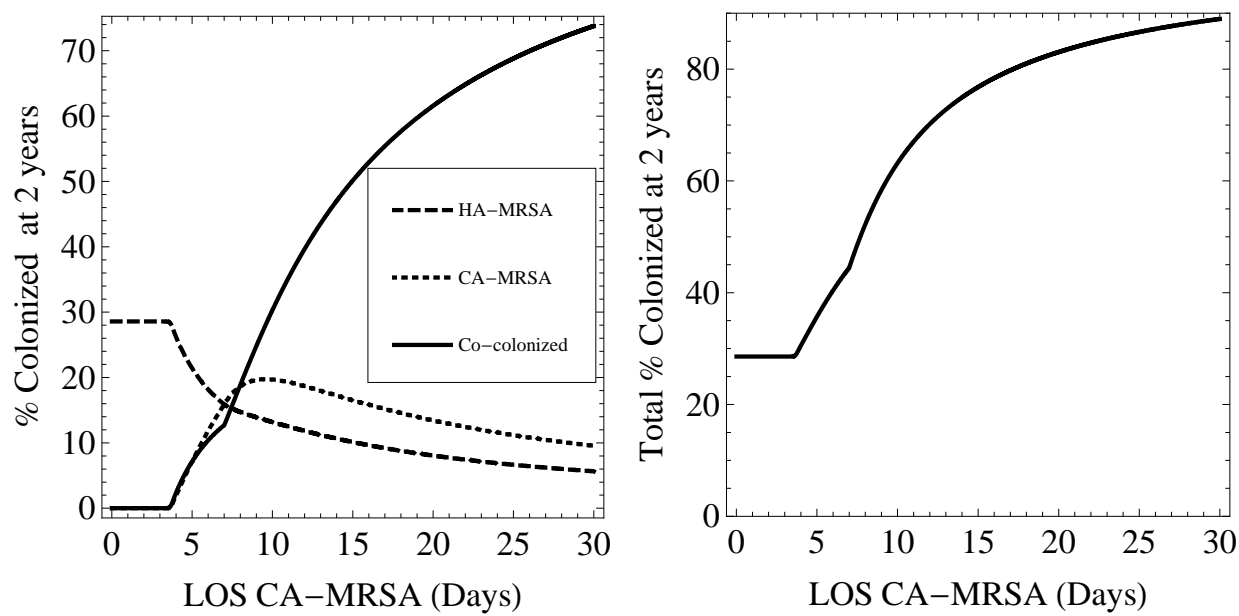}
\caption{Left - percentage of patients colonized with HA-MRSA (dashed), CA-MRSA (dotted), and both (solid) after 2 years, as the length of stay of patients colonized with CA-MRSA is increased. Right - the total percentage of patients colonized with MRSA after 2 years as LOS is increased. LOS of patients with HA-MRSA is 7 days.  LOS of patients colonized with both strains equals the greater of the LOS of patients colonized with CA-MRSA or HA-MRSA.}
\label{fig5}
\end{figure}

Increasing the rate of admission of patients colonized with CA-MRSA or increasing their LOS results in an increase in the co-colonized state (figures \ref{fig4} and \ref{fig5}).  LOS has a substantially greater impact on the prevalence of co-colonization compared to rate of admission. Even a minimal increase in LOS from the baseline value of 5 days to 8 days leads to the majority of colonized patients represented by the co-colonized state.  The predominance of the co-colonization state when LOS increases is explained by the increased opportunity for HA-MRSA colonized patients to become colonized with CA-MRSA and become colonized with both strains.

\subsection*{Simulation 3: interventions to prevent transmission}
\begin{figure}
\includegraphics[width=0.75\textwidth]{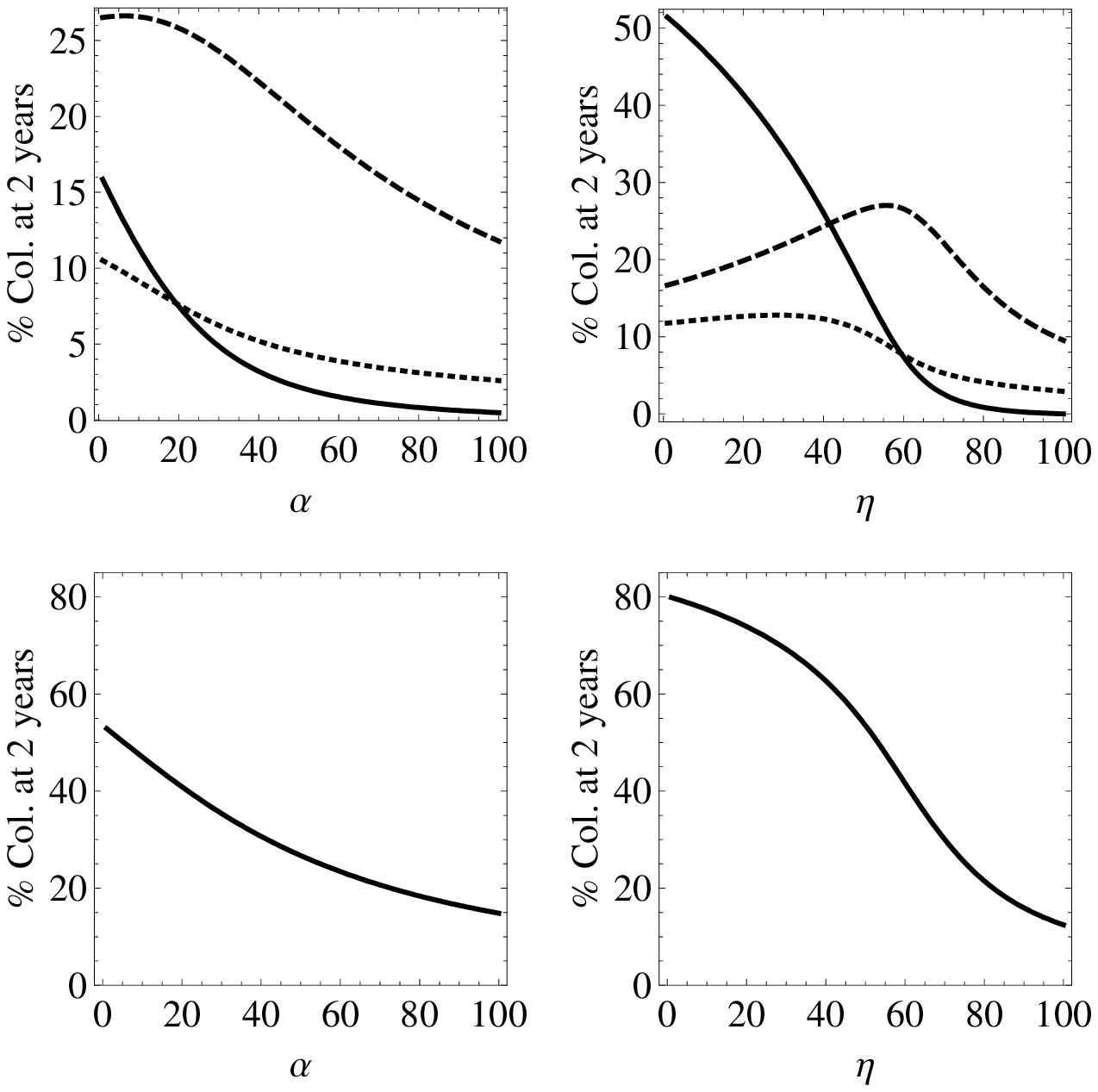}
\caption{Top - percentage of patients colonized with HA-MRSA (dashed), CA-MRSA (dotted), and both (solid) after 2 years, versus two interventions: left - decolonization ($\alpha$) and right - hand-hygiene compliance ($\eta$).  Bottom - sum of percentages of patients colonized with either or both strains of MRSA. $100\lambda_C =3\%$, $100\lambda_H = 7\%$, $100\lambda_B = 0\%$.}
\label{fig6}
\end{figure}

The effect of two standard interventions aimed at preventing the transmission of MRSA, improving compliance with hand-hygiene and maximizing the efficacy of decolonization of patients with MRSA, were evaluated in simulations which included an influx of patients harboring HA-MRSA or CA-MRSA.  Both interventions decrease the percentage of colonized patients in all three states. Compared to decolonization, improving hand-hygiene has the greater impact with even small increases in compliance having a substantial effect in the overall prevalence of MRSA. Since there is a constant influx of colonized patients into the hospital setting, CA- and HA-MRSA are never eradicated from the hospital even at 100\% hand-hygiene compliance or 100\% decolonization efficacy.  In contrast, the absence of an influx of patients who are co-colonized into the hospital explains the extinction of the B state when these two interventions are at 100\% compliance or efficacy.  As hand-hygiene compliance increases, the total percentage of patients colonized decreases monotonically.  However, simulations show that after 2 years, the percentage of patients who only have HA-MRSA increases until hand-hygiene reaches about 45\% (figure \ref{fig6}).

The explanation for this finding is as follows. Susceptible patients move to the single colonized state with a rate $(1-\eta)\beta$ and from the single colonized state to the both state with a rate $(1-\eta)\beta$.  Therefore there is a quadratic effect of hygiene on transmission to the co-colonized state. When hand-hygiene compliance is low (and the co-colonized compartment is large), the quadratic effect of increasing hand-hygiene compliance, causes a rapid reduction of the co-colonized compartment, and increases the population of susceptible patients in the equilibrium. Transmission to the single colonization compartments ($C$ and $H$) is dependent both on $\eta$ and $S$ by the term $\frac{(1-\eta)\beta}{N}S$.  Until hand-hygiene reaches about 45\%, the rise in $S$ due to the quadratic effect on the co-colonization state is stronger than the reduction in transmission due to increasing $\eta$.  Therefore, the sizes of the $C$ and $H$ compartments increase.

\subsection*{Simulations assuming greater transmission potential for CA-MRSA compared to HA-MRSA}
The overall results of simulations with $\beta_C =1.33\beta_H$  were similar to those with $\beta_C = \beta_H$ except for a greater and more rapid increase in CA-MRSA and co-colonized patients (data not shown).

\section*{Discussion}
The transmission dynamics of MRSA in the hospital setting are complex and require the analysis of numerous interrelated and dynamic factors. The emergence of CA-MRSA and its invasion into the hospital setting has led to further complexities in understanding not only the spread of MRSA, but also the selection of effective antimicrobial therapies and preventive strategies.  Since epidemiological studies cannot fully address these complexities, a mathematical model was developed to specifically quantify the emergence of individuals co-colonized with both CA-MRSA and HA-MRSA strains.

	The deterministic model shows that over time, there will be no competitive exclusion of either strain but that both CA-MRSA and HA-MRSA will co-exist in the hospital setting. The model also shows that individuals co-colonized with both strains will increase in prevalence over time and will predominate over individuals who are colonized with either CA-MRSA or HA-MRSA.  These findings have important implications.  First, clinical cultures will usually identify only one strain, either CA-MRSA or HA-MRSA.  Mathematical models have shown that 20 colonies per patient need to be sampled to reliably estimate the occurrence of multiple strains \cite{17}.  Since sampling of multiple colonies is not feasible, co-colonization and polyclonal infections will therefore not be routinely detected.  The different antimicrobial susceptibility profiles of CA-MRSA and HA-MRSA and identification of only one strain may therefore lead to incorrect antimicrobial therapy.  The emergence of multidrug-resistant strains of CA-MRSA, with susceptibility profiles similar to HA-MRSA may however minimize difficulties in selection of the appropriate antimicrobial among co-colonized patients \cite{18}.

	Factors which increase the reservoir of CA-MRSA in the hospital setting were shown to have substantial effects on the magnitude of the co-colonized patient population.  Increasing the influx of patients harboring CA-MRSA into the hospital and increasing their LOS resulted in a rapid increase in the number of co-colonized patients.  Previous mathematical models have also shown that these two factors are central to the dissemination of antimicrobial-resistant bacteria \cite{7,19}.  Our model shows that even small increases in LOS of only few days from a conservative baseline estimate of 5 days among CA-MRSA patients resulted in the rapid predominance of patients colonized with both strains.  CA-MRSA is associated with severe infections and several studies have documented that CA-MRSA has become the predominant MRSA strain implicated in the great majority of nosocomial blood stream infections and surgical site infections \cite{4,20,21}.  These nosocomial infections would therefore substantially extend the LOS of patients harboring CA-MRSA.

	Our model revealed the effects of two interventions targeting the prevention of MRSA spread: improving compliance with hand-hygiene and decolonization strategies.  Both interventions decreased the prevalence of HA-MRSA, CA-MRSA and co-colonization.  As shown in previous models, improving hand-hygiene compliance had the greatest effect with only small improvements in compliance \cite{7}.

	Several assumptions were made to simplify the model.  First, antibiotic pressure and its effect on the transmission dynamics of CA-MRSA and emergence of co-colonization were not assessed.  Given the different susceptibility profiles between HA-MRSA and CA-MRSA, selective antibiotic pressure may alter the transmission dynamics between HA-MRSA and CA-MRSA.  Second, environmental contamination was not included since there is a paucity of data regarding differences in contamination of inanimate surfaces between CA-MRSA and HA-MRSA.  Future models will need to include these factors.  Our main model assumed that the likelihood of transmission between CA-MRSA and HA-MRSA from HCW to patient and vice versa was equal.  In vitro studies have shown that the growth rate of certain CA-MRSA strains is faster compared to HA-MRSA strains suggesting that CA-MRSA may have an advantage towards colonization and therefore greater transmissibility \cite{11,12}.  Model simulations using greater transmission parameters for CA-MRSA showed similar conclusions to the baseline model except for more rapid dynamics of CA-MRSA spread and emergence of co-colonization.  Lastly, a deterministic model was used which compartmentalizes patients into homogeneous groups and therefore individual-level behavior was not addressed.  Although stochastic individual-based models can simulate the heterogeneity of patients and HCW interactions, the increase in behavioral details provided by these models may result in greater difficulty in interpretation of findings.

	Colonization with different strains and species of bacteria is common \cite{9}.  This study quantified the emergence and spread of co-colonization with both CA-MRSA and HA-MRSA. The study demonstrated that the expanding community reservoir of CA-MRSA resulting in an increase influx of CA-MRSA patients into the hospital setting coupled with prolonged LOS associated with severe CA-MRSA infections will rapidly lead to a predominance of patients who are colonized with both strains.  The impact of these findings on patient outcomes and the potential for transfer of genetic information between these strains will require ongoing evaluation.


\newpage
\acknowledgments
\section*{Acknowledgments}
This work was supported in part by the joint DMS/NIGMS Initiative through the National Institute of Health Grant R01GM083607 (EMCD, GFW, JP).

Potential Conflicts of interest: EMCD, GFW, JP no conflicts.



\appendix

\section*{Appendix}

A reduced version of D'Agata et al \cite{7} was developed, which exhibits the same qualitative properties as the original, but which lacks the infective states.  The patient population is split into three compartments: $S$ - susceptible patients, $C$ - patients colonized with CA-MRSA and $H$ - patients colonized with HA-MRSA. We added the assumption that the total number of patients was conserved at size $N = 400$, allowing us to reduce by one dimension to the compartments $C$ and $H$, by letting $S=N-C-H-B$. The model is analyzed without this assumption in an accompanying paper, and the results are qualitatively the same.

We then extended the model, allowing patients to be concurrently co-colonized with CA-MRSA and HA-MRSA, which adds a third (or fourth if there is no conservation) compartment, $B$ - patients colonized with "B"oth CA-MRSA and HA-MRSA. The rate of change of the size of the compartments due to the transmission of MRSA in the hospital is then described by the following system of nonlinear ordinary differential equations:

\begin{align}
\label{eq:fullsystem11}
\frac{dC}{dt} &= (\delta_S S + \delta_C C + \delta_H H)\lambda_C + \frac{(1-\eta)\beta_C}{N} S (C + B) - \frac{(1-\eta)\beta_{CH}}{N}C(H + B) - (\delta_C + \alpha_C)C\\
\frac{dH}{dt} &= (\delta_S S + \delta_C C + \delta_H H)\lambda_H + \frac{(1-\eta)\beta_H}{N} S (H + B) - \frac{(1-\eta)\beta_{CH}}{N}H(C + B) - (\delta_H + \alpha_H)H\\
\frac{dB}{dt} &= (\delta_S S + \delta_C C + \delta_H H)\lambda_B + \frac{(1-\eta)\beta_{CH}}{N}C(H + B) + \frac{(1-\eta)\beta_{HC}}{N}H(C + B) - (\delta_B + \alpha_B)B,
\label{eq:fullsystem14}\end{align}
with $S=N-C-H-B$. Parameter explanations and values are given in Table \ref{table1}.

Assuming that all patients enter the hospital susceptible ($\lambda_C = \lambda_H = \lambda_B = 0$), there exists a disease free equilibrium (DFE) of the system, where all patients are susceptible, $S=N$ and $(C, H, B) = (0, 0, 0)$. By linearizing the system and evaluating the eigenvalues of the Jacobian matrix, we find that the DFE exists for all parameters and is locally asymptotically stable if $max\{R0^H, R0^C\}<1$, where $R0^H = \frac{(1-\eta)\beta_H}{\alpha_H + \delta_H}$ is the basic reproduction number for HA-MRSA and $R0^C = \frac{(1-\eta)\beta_C}{\alpha_C + \delta_C}$ is the basic reproduction number for CA-MRSA. This means that if $max\{R0^H, R0^C\}<1$, both strains of MRSA will be extinguished over time.

 In addition to the DFE, there are two other analytically known equilibria, $E_H$ and $E_C$, which describe states where one disease is endemic while the other absent. When $E_H$ is stable, HA-MRSA will be endemic in the hospital and CA-MRSA will be extinguished over time.  Therefore, the size of the compartments $S$ and $H$ will be positive and the size of compartments $C$ and $B$ will be zero. Symmetrically, when $E_C$ is stable, CA-MRSA will be endemic and HA-MRSA will be extinguished over time.

A fourth equilibrium, in which the size of every compartment is positive, does not have a known analytic form but is consistently found in numerical simulations.

The invasion reproduction number \cite{22} $I0$ is a threshold parameter which determines the stability of the equilibria $E_H$ and $E_C$.  When $I0^{E_H} > 1$, $E_H$ is unstable, CA-MRSA invades, or becomes endemic, and both strains become prevalent in the hospital over time. Conversely, if $I0^{E_H} < 1$, $E_H$ is stable and only HA-MRSA will be endemic in the hospital over time.  Symmetrically, when $I0^{E_C} > 1$, $E_C$ is unstable and both strains become endemic over time. Conversely, if $I0^{E_C} < 1$, $E_C$ is stable and only CA-MRSA will be endemic in the hospital.

\begin{equation}
I0^{E_H} = \frac{R0^C}{R0^H}\left(\frac{\frac{(1-\eta)\beta_{CH}}{(\alpha_B+\delta_B)} \left(1-\frac{1}{R0^H}\right) + 1}{\frac{(1-\eta)\beta_{CH}}{(\alpha_C+\delta_C)} \left(1-\frac{1}{R0^H}\right) + 1}\right) + \frac{(1-\eta)\beta_{HC}}{\alpha_B+\delta_B} \left(1-\frac{1}{R0^H}\right).
\end{equation}

A symmetric form is found for $I0^{E_C}$, since the model is symmetric in $C$ and $H$.

Assuming $\beta_C = \beta_H = \beta_{CH} = \beta_{HC}$ and $\alpha_C = \alpha_H = \alpha_B$, as we do in this paper,
\begin{equation*}
I0^{E_H} = \frac{1}{1-\frac{1}{R0^H} + \frac{1}{R0^C}} + R0^H - 1
\end{equation*}

When $R0^H> 1$ and $R0^C > 1$, $I0^{E_H} > 1$.


Symmetric results hold for $I0^{E_C}$. Therefore, neither endemic equilibrium is stable when $R0^H> 1$ and $R0^C > 1$, and there is never competitive exclusion. If patients continually enter the hospital colonized with MRSA, then it can never be completely extinguished. In an accompanying paper, the model is more thoroughly investigated mathematically.




\clearpage
\begin{table}
\caption{Parameter estimates for the transmission dynamics of community-acquired and hospital-acquired methicillin-resistant \textit{Staphylococcus aureus} colonization (CA-MRSA and HA-MRSA).}

\

\footnotesize{
\begin{tabular}{| l | l | l | l |}
\hline
\textbf{Parameter} & \textbf{Symbol} & \textbf{Baseline Value} & \textbf{Source} \\
\hline
\hline
Total number of patients & N & 400 & \\
\hline
\underline{Percent of admissions per day} & & &\\
Colonized CA-MRSA & 100 $\lambda_C$ & 3 & 13,14\\
Colonized HA-MRSA & 100 $\lambda_H$ & 7 & BI, 13, 14\\
\hline
\underline{Length of stay} & & & \\
Susceptible & $1/\delta_S$ & 5 days & BI\\
Colonized CA-MRSA & $1/\delta_C$ & 5 days & BI\\
Colonized HA-MRSA & $1/\delta_H$ & 7 days & 5\\
Co-colonized & $1/\delta_B$ & 7 days &\\
\hline
Hand-hygiene compliance efficacy (as \%)& 100 $\eta$ & 50\% &\\
\hline
\underline{Transmission rate per susceptible patient to} & & & \\
Colonized CA-MRSA per colonized CA-MRSA & $\beta_C$ & 0.4 per day & 11,12 \\
Colonized HA-MRSA per colonized HA-MRSA & $\beta_H$ & 0.4 per day & 11,12 \\
\hline
\underline{Transmission rate per patient colonized with CA-MRSA to} & & & \\
Co-colonized per colonized CA-MRSA & $\beta_{CH}$ & 0.4 per day & 11,12 \\
\hline
\underline{Transmission rate per patient colonized with HA-MRSA to} & & & \\
Co-colonized per colonized HA-MRSA & $\beta_{HC}$ & 0.4 per day & 11,12 \\
\hline
\underline{Decolonization rate per colonized patient} & & &\\
 \underline{per day per length of stay (as \%)} & & &\\
CA-MRSA & 100 $\alpha_C$ & 0\% & 15,16 \\
HA-MRSA & 100 $\alpha_H$ & 0\% & \\
Co-Colonized & 100 $\alpha_B$ & 0\% & \\
\hline
\end{tabular}
BI: data obtained from the Beth Israel Deaconess Medical Center
}
\label{table1}

\end{table} 
\clearpage



\clearpage


\begin{thebibliography}{}
\bibitem{1} Klevens RM, Morrison MA, Nadle J et al. Invasive methicillin-resistant \textit{Staphylococcus aureus} infections in the United States. JAMA \textbf{2007}; 298:1763-71
\bibitem{2} Herold BC, Immergluck LC, Maranan LC  et al. Community-acquired methicillin-resistant \textit{Staphylococcus aureus} in children with no identified predisposing risk. JAMA \textbf{1998}; 279:593-98.
\bibitem{3} Tristan A, Bes Meugnier H et al. Global Distribution of Panton-Valentine Leukocidin-positive Methicillin-resistant \textit{Staphylococcus aureus}, \textbf{2006}.  Emerg Infect Dis 2007;13:594-600.
\bibitem{4} Patel M, Waites KB, Hoesley CJ, Stamm AM, Canupp KC, Moser SA. Emergence of USA 300 MRSA in a tertiary medical centre: implications for epidemiological studies. J Hosp Infect \textbf{2008};68:208-13.
\bibitem{5} Davis SL, Rybak MJ, Amjad M, Kaatz GW, McKinnon PS. Characteristics of patients with healthcare-associated infection due of SCCmec Type IV methcillin-resistant \textit{Staphylococcus aureus}. Infect Control Hosp Epidemiol \textbf{2006};27:1025-31.
\bibitem{6} Popovich KJ, Weinstein RA, Hota B. Are community-associated methicillin-resistant \textit{Staphylococcus aureus} (MRSA) strains replacing traditional nosocomial MRSA strains? Clin Infect Dis. \textbf{2008};46:795-8.
\bibitem{7} D'Agata EMC, Webb GF, Horn MA, Moellering RC, Ruan S. Modeling the invasion of community-acquired methicillin-resistant \textit{Staphylococcus aureus} into hospitals. Clin Infect Dis \textbf{2009}:48; 274-84.
\bibitem{8} Cespedes C, Said-Salim B, Miller M, Lo SH, Kreiswirth BN, Gordon RJ, Vavagiakis P, Klein RS, Lowy FD. The clonality of \textit{Staphylococcus aureus} nasal carriage. J Infect Dis. \textbf{2005};191:444-52.
\bibitem{9} Lautenbach E, Tolomeo P, Black N, Maslow JN. Risk factors for fecal colonization with multiple distinct strains of \textit{Escherichia coli} among long-term care facility residents. Infect Control Hosp Epidemiol. \textbf{2009};30:491-3.
\bibitem{10} Li M, Diep BA, Villaruz AE, Braughton KR, Jiang X, DeLeo FR, Chambers HF, Lu Y, Otto M. Evolution of virulence in epidemic community-associated methicillin-resistant \textit{Staphylococcus aureus}. Proc Natl Acad Sci U S A. \textbf{2009};106:5883-8.
\bibitem{11} Okuma K, Iwakawa K, Turnidge JD et al. Dissemination of New Methicillin-Resistant \textit{Staphylococcus aureus} Clones in the Community. J Clin Microbiol \textbf{2002};40:4289-94.
\bibitem{12} Baba T, Takeuchi F, Kuroda M et al. Genome and virulence determinants of high virulence community-acquired MRSA. Lancet 2002;359:1819-27.
\bibitem{13} Hidron AI, Kourbatova EV, Halvosa JS et al. Risk factors for colonization with methicillin-resistant \textit{Staphylococcus aureus} (MRSA) in patients admitted to an urban hospital: emergence of community-associated MRSA nasal carriage. Clin Infect Dis \textbf{2005};15:159-66.
\bibitem{14} Jarvis WR, Schlosser J, Chinn RY, Tweeten S, Jackson M. National prevalence of methicillin-resistant \textit{Staphylococcus aureus} in inpatients at US health care facilities, \textbf{2006}. Am J Infect Control 2007;35:631-37
\bibitem{15} Diekema DJ, Climo M. Preventing MRSA infections: finding it is not enough. JAMA. \textbf{2008};299:1190-92.
\bibitem{16} Wertheim HF, Melles DC, Vos MC  et al. Effect of mupirocin treatment on nasal, pharyngeal, and perineal carriage of \textit{Staphylococcus aureus} in healthy adults. Antimicrob Agents Chemother. \textbf{2005};49:1465-67.
\bibitem{17} Coen PG, WIlks M, Dall'Antonia, Millar M. Detection of multiple-strain carriers: the value of re-sampling. J Theor Biol \textbf{2006}:240;98-103.
\bibitem{18} Diep BA, Chambers HF, Graber CJ, Szumowski JD, Miller LG, Han LL, Chen JH et al. Emergence of multidrug-resistant, community-associated methicillin-resistant \textit{Staphylococcus aureus} clone USA300 in men who have sex with men. Ann Intern Med \textbf{2008};148:249-57.
\bibitem{19} D'Agata EMC, Webb G. Horn M. A mathematical model quantifying the impact of antibiotic exposure and other interventions on the endemic prevalence of vancomycin-resistant enterococci. J Infect Dis \textbf{2005};192:2004-11.
\bibitem{20} Miller LG, Perdreau-Remington F, Rieg G et al. Necrotizing fasciitis caused by community-associated methicillin-resistant \textit{Staphylococcus aureus} in Los Angeles. New Engl J Med \textbf{2005};352:1445-53.
\bibitem{21} Seybold U, Kourbatova EV, Johnson JG et al. Emergence of community-associated methicillin-resistant \textit{Staphylococcus aureus} USA 300 genotype as a major cause of health care-associated blood stream infections. Clin Infect Dis \textbf{2006};42;647-66.
\bibitem{22} Xiridou M, Borkent-Raven B, Hulshof J, Wallinga J. How Hepatitis D Virus Can Hinder the Control of Hepatitis B Virus. PLoS ONE. \textbf{2009}; 4:e5247.
\end{thebibliography}
\end{document}